# Matters Arising

# The ALPHA-g experiment, the hunt for gravitational dipoles and the quantum vacuum as the source of gravity in the Universe

Dragan Slavkov Hajdukovic
INFI, Cetinje, Montenegro
dragan.hajdukovic@gmail.com , dragan.hajdukovic@alumni.cern

**Abstract**
More recently, the ALPHA-g experiment has shown that *antiatoms* fall in the Earth's gravitational field like ordinary atoms, with the non-negligible possibility that antiatoms fall with a slightly lower acceleration. A possible lower acceleration of antiatoms (which could be revealed with two orders of magnitude higher precision measurements) would be an epoch-making discovery and a window to a new physics, suggesting among other things that quarks and antiquarks have gravitational charges of opposite sign. If gravitational charges of opposite sign exist, it implies the existence of virtual gravitational dipoles in the quantum vacuum, and opens up the possibility of including the quantum vacuum as an important source of gravity in the Universe.

Physics would be more poetic if antiatoms fell up instead of down, but the incredibly sophisticated, difficult and fascinating experiment[1] of the ALPHA-g collaboration tells us that this is not the case. For the first time in human history, we know with certainty that atoms and antiatoms fall in the same direction. More precisely, the best fit to the ALPHA-g measurements gives

$$a_{\bar{H}} = [0.75 \pm 0.13(\text{statistical} + \text{ systematic}) \pm 0.16 \text{ (simulation)}]g$$

for the local acceleration of antihydrogen towards the Earth. The significance of this breakthrough was immediately recognised[2,3].

Although the above result gives the impression that antiatoms fall with a lower acceleration than atoms, it is premature to make such a claim (since the precision of the current ALPHA-g experiment is no better than 20% of g). However it is illuminating to give a few probabilities based on normal distributions with (according to ALPHA-g) mean value $a_{\bar{H}} = 0.75g$ and standard deviation *0.3g*.

$$P(a_{\bar{H}} < g) \approx 0.8$$
$$P(a_{\bar{H}} < 0.9g) \approx 0.7$$
$$P(a_{\bar{H}} < 0) \approx 0.006$$
$$P(a_{\bar{H}} < -0.4g) \approx 0.0001$$

As we can see, according to the current ALPHA-g measurements, the probability that antiatoms fall with a lower acceleration is about 0.8; there remains a probability of 0.2 that antiatoms fall with a higher acceleration than atoms. Even an order of magnitude increase in the precision of the measurements can result in $P(a_{\bar{H}} < g) \approx 0.999$. By the way, note that the ALPHA-g experiment rules out the possibility that atoms and antiatoms have opposite accelerations of the same magnitude (probability of it is about 10[-15]).

In short, the ALPHA-g experiment only determined the sign of the acceleration and gave a rough approximation of the magnitude of the acceleration. Both, atoms and antiatoms fall towards the Earth, but it remains possible that the magnitude of the acceleration is different, i.e. that antiatoms fall with a smaller acceleration $a_{\bar{H}} < a_H = g$ (For an illuminating study of this possibility, see a recent paper[4] by Scott Menary). In other words, in the gravitational field of the Earth, atoms and antiatoms have a positive gravitational mass (gravitational charge), but the gravitational charge of antiatoms ($m_{g\bar{H}}$) could be smaller, $m_{g\bar{H}} < m_{gH}$. (Note that on an anti-Earth, assuming CPT symmetry, the above conclusions would be exactly the opposite, $a_{\bar{H}} > a_H$ and $m_{g\bar{H}} > m_{gH}$).

Hopefully, already within this decade the gravitational acceleration of antiatoms will be measured to a much higher precision of the order of $10^{-2}$ or $10^{-3}$ (up to $10^{-6}$, in the best case[5]). Any difference in the gravitational acceleration of atoms and antiatoms would be an epoch-making discovery and a window to a new physics.

A major question is what ALPHA-g tells us about the gravitational interactions of the fundamental particles of the Standard Model of particles and fields, i.e. about quarks and antiquarks, leptons and antileptons, gauge bosons (photons, eight gluons, $Z^0$ and $Z^{\pm}$) and the Higgs boson. The somewhat surprising answer is nothing.

The most important fact is that the ALPHA-g measurements, do not allow us to say whether the gravitational interactions between the fundamental particles are always attractive or in some cases repulsive. While there is an established gravitational attraction between a proton and an antiproton (or an Earth and an anti-Earth), it remains possible for a quark and an antiquark, or a lepton and an antilepton, to repel each other.

There is a significant difference between a fundamental structureless particle and a bound state (which may look like a particle but has structure). In the Standard Model, fundamental particles are structureless, so quarks and leptons are just matter, while antiquarks and antileptons are just antimatter. In contrast, particles with structure (such as protons and antiprotons) are made up of both matter and antimatter. The proton is a relativistic bound state with short-lived virtual quark-antiquark pairs (hence some of the proton's mass is antimatter), gluons and three valence quarks ($uud$). If the three valence quarks ($uud$) are replaced by three valence antiquarks ($\bar{u}\bar{u}\bar{d}$), the proton becomes an antiproton; while this does not change the mass, the antiproton contains more antimatter than the proton. Consequently, if antiquarks eventually have a negative gravitational charge, the gravitational charge of the antiproton would be less than its mass, and the acceleration of the antiproton would be less than the acceleration of the proton in the Earth's gravitational field.

Assuming that antiquarks have a negative gravitational charge, it is easy to estimate the upper limit of the gravitational acceleration ($a_{\bar{H}}$) of antihydrogen. The mass of three valence quarks ($uud$) and three valence antiquarks ($\bar{u}\bar{u}\bar{d}$) is equal to about 0.01 of the total mass of the proton or antiproton. So if 0.01 of the proton's positive gravitational charge is replaced by a negative gravitational charge, the antiproton's gravitational charge remains positive, but is reduced to 0.98 of the proton's gravitational charge (so the antiproton's gravitational acceleration would be less than that of a proton). Since the valence antiquarks in an antiproton are relativistic (and therefore have more energy than non-relativistic quarks), the inequality $a_{\bar{H}} < 0.98g$ defines the upper limit for $a_{\bar{H}}$.

In short, if the forthcoming experiments with higher precision reveal an anomalous acceleration of antiatoms ($a_{\bar{H}} < 0.98g$), this would be a serious indication that quarks and antiquarks have the gravitational charges of the opposite sign. It is therefore an open question whether there is a negative gravitational charge of antiquarks, which is obscured by the composite structure of antiprotons and the current low precision of ALPHA-g measurements. Metaphorically speaking, ALPHA-g showed that there is gravitational attraction between an apple and an anti-apple, but it remains possible that there is gravitational repulsion between a quark and an antiquark, or an electron and an antielectron. Only by increasing the precision of the ALPHA-g measurements can we gain some insight into the gravitational interactions of the antiquarks and the possible existence of gravitational repulsion between quarks and antiquarks.

In principle, ALPHA-g and two competing experiments (AEGIS[6] and GBAR[7]) are dedicated to the quark sector of the Standard Model. Fortunately, complementary experiments are planned to measure the gravitational acceleration of positronium[8] (a bound state of electron and antielectron) and muonium[9] (a bound state of electron and antimuon). If antileptons had a negative gravitational charge, the measured acceleration of positronium would be zero (because positronium has equal amounts of matter and antimatter, so its gravitational charge is zero). Unlike positronium, muonium would be repelled by the Earth and fall upwards (because muonium, with an antimuon 200 times more massive than the electron, is almost pure antimatter and its gravitational charge is negative).

Regardless of how gravitation works between fundamental particles, the ALPHA-g result, which confirms the gravitational attraction between matter and antimatter nuclei and atoms, has already ruled out some alternative cosmological ideas, including the so-called Dirac-Milne cosmology[10] (which advocates a Universe with equal amounts of matter and antimatter distributed in different regions separated by gravitational repulsion).

At first sight, the results of ALPHA-g rule out any influence of antimatter on the evolution of the present-day Universe. However, there is a hidden possibility for antimatter to have a major impact on the Universe; this possibility is hidden in the quantum vacuum, which is a plausible source of what we call dark energy.

In fact, the ALPHA-g experiment (together with all other antimatter gravity experiments) is the beginning of the search for gravitational dipoles, comparable to the search for magnetic monopoles. As is well known, there are magnetic dipoles without any empirical evidence for magnetic monopoles; in the case of gravity it is exactly the opposite, there are gravitational monopoles without any evidence for gravitational dipoles. However, at the fundamental particles level, if there is gravitational repulsion between quarks and antiquarks, then all quark-antiquark pairs are gravitational dipoles; similarly, if there is gravitational repulsion between leptons and antileptons, then all lepton-antilepton pairs (such as electron-antielectron pairs) are gravitational dipoles. This opens up the far-reaching fundamental possibility that quantum-vacuum fluctuations (at least the part of the fluctuations that can be regarded as virtual particle-antiparticle pairs) are gravitational dipoles.

It is worth noting that within the framework of General Relativity (assuming the validity of CPT symmetry), gravitational dipoles associated with antimatter[11] can exist, while within the framework of extended General Relativity (independent of antimatter), gravitational dipoles associated with spin and torsion[12] can exist.

Physics is in a major crisis, and we shouldn't dismiss any possibility without serious study. One such possibility is the hypothesis that quantum-vacuum fluctuations are, by their very nature, virtual gravitational dipoles. Rough preliminary studies[13, 14] of the consequences of this hypothesis suggest that the quantum vacuum, enriched with virtual gravitational dipoles, could be a common explanation for phenomena usually attributed to hypothetical dark matter and dark energy.

In conclusion, antimatter-gravity experiments are an important emerging field of research which, with a significant increase in the precision of the measurements, has a surprising potential to provide important insights into the gravitational interactions of the fundamental particles of the Standard Model of Particles and Fields, and thus into the gravitational properties of the quantum vacuum, which may ultimately play a crucial role in the evolution of the Universe.